# Thermal acoustic particle velocity sensor with structured microwires

Ruby Jindal[a,b, *], Sushil Kumar Singh[b], Ravibabu Mulaveesala[a] and Jolly Xavier[a, *]

[a] *SeNSE, Indian Institute of Technology, Hauz Khas, New Delhi-110016, India*
[b] *SSPL, Lucknow Road, Timarpur, New Delhi – 110054, India*

E-mail: jxavier@sense.iitd.ac.in
idz228533@sense.iitd.ac.in

**Abstract**
Estimating the spatial distribution of the acoustic field is essential in communication, medical imaging and other varied industrial processes. Acoustic particle velocity sensor plays a key role in providing directional information of the sound field. We present a rigorous 3D numerical finite element analysis of MEMS based structurally modified thermal particle velocity sensor. The impact of diverse morphological structural optimization in comparison to the conventional straight wire structure is rigorously studied and analyzed in order to achieve maximum temperature deviation and, thereby, the sensitivity of the device primarily for low frequency applications.

## 1. Introduction

Acoustic sensing detects, measures, and analyses sound signals in various environments. It focuses on capturing and interpreting these acoustic signals. The two key parameters, acoustic pressure, which provides omnidirectional scalar information, and acoustic particle velocity, which offers vector-based directional data, describe the complete sound field [1]. Conventional microphones make the system bulky, which is also not handy for low-frequency applications due to the spatial sampling criteria[2].Directional acoustic sensing is required for various applications, such as warfare and battlefield acoustics [3]. Therefore, the need for a miniaturized acoustic vector sensor arises. An acoustic vector sensor (AVS) combines four-channel sensors comprising of one non-directional sound pressure sensor and three mutually perpendicular acoustic particle velocity sensor (APVS), individually sensitive in single direction. AVS measures acoustic intensity of the sound field. It is a vector measure possessing both magnitude and direction[4]. In the last three decades, the innovation of sensors for measuring particle velocity has shown a paradigm shift in acoustics. They enable to measure the particle velocity of incoming acoustic signal without any intermediate conversion. It contains both amplitude and phase information [5]. Particle velocity sensor is one of its kinds, utilizes thermal sensing principle to measure the particle velocity of a signal for airborne applications. APVS have gained a lot of importance in every field, whether deterrence or commercial applications [6]. AVS is used to detect a particular acoustic phenomenon[7]. Acoustic signal detection and processing are implemented in various real-life applications such as underwater applications[8], Aircraft detection[9][10], Non-destructive testing of materials[11], detecting the sound level of the source due to HVAC(Heating, Ventilation and Air Conditioning)[12], mapping sound intensity in car/cabin interiors[13], wind measurements[14], monitoring purposes[15], UAV(Unmanned Air Vehicles) detection[16], urban noise monitoring[17], sound source localization, engine fault detection[18], mapping of acoustic intensity in the vicinity of human head[19], battlefield acoustics[3] and many more. In addition to developing algorithms for different applications, it is also essential to have studied the behaviour of particle velocity sensor. The cantilever type, bridge type, sensor heater sensor, and through the wafer type are the four structures given by particle velocity sensor[5]. Among these, bridge-type structure is most commonly used for commercial purposes due to their better mechanical stability and low power consumption[6].

     The bridge-type structure consists of two parallel heated wires strengthened via dielectric beams of similar dimensions. The principles of hot wire anemometry and mass flow sensor make the working principle of Particle velocity sensor. The wires are heated via constant current/constant temperature to attain optimum temperature. The perpendicular interaction of acoustic signals with the heated wires generates the temperature gradient, which is directly related to the particle velocity[6][20] . Since it may be used as a mass flow sensor and is susceptible to low-frequency sound signals, the sensor serves as an add-on microphone. Further research and development of the device resulted in introducing a commercial acoustic probe based on the same APVS sensor[21].The directivity pattern of particle velocity sensor is similar to 'figure of eight'. Due of its frequency independence, the directivity effect is capable of being used in a variety of real-world operational scenarios with background noise. The operating frequency encompasses the entire frequency range[22].

Over the years, various researchers have discussed the behaviour of Thermal acoustic particle velocity sensors (TAPVS). A four-wire particle velocity sensor pairs wires are in such a way that their sensitivity directions are mutually perpendicular and provide at least two uncorrelated signals in each particular direction, which reduces the contribution of overall uncorrelated noise[23]. The incorporation of a third wire in the centre serves as an auxiliary heating element, improving the sensitivity and efficiency of the device. The use of only two wires limited its performance. The manually assembled sensor performs better than the four-wire APVS[5].

Another modification of four parallel wire results in a two-dimensional particle velocity sensor and used as a 3-dimensional sensor when combined. It provides better sensitivity and improved noise performance due to the interaction between parallel wire pairs acting as heaters to one another. However, the device faces difficulty in assembly due to manual alignment, which results in imperfect/false results[24].

A revised four-wire sensor configuration exhibits a reduced noise floor compared to its predecessor and is capable of delivering two-dimensional data pertaining to the field[25]. A 2-dimensional particle velocity sensor configuration in which wires crosswise minimizes the flow disturbance through freely suspended wires. It also enhances the sensitivity and reduces the overall power consumption of the device. However, it has some limitations, including a minor mismatch of about 0.4 dB between channels, attributed to inaccuracies in mask printing, and a 10% mismatch in wire resistance between the two directions. Also, the package gain was less than that of the simulated value[26][27]. A five-wire APVS with 3D directionality can detect low airflow and particle velocity measurements. It shows good sensitivity and directionality with minimized self-heating effects. However, there are challenges in measuring low gas speeds, and it is more prone to environmental noise interference[28][29]. A one-dimensional, cost-effective particle velocity sensor employs a commercial post-CMOS MEMS process. There is a noticeable rise in heat loss through the arms of the membrane attributed to the design, resulting in an equivalent noise level that exceeds that of sensors produced using fully dedicated technologies[30]. We present a complete rigorous heating and thermos-viscous acoustics with a non-linear coupling of material properties for a comprehensive numerical analysis of SMTAPVS using three dimension simulations. A complete methodology is utilized, encompassing joule heating and thermos-viscous acoustics. The proposed designs are useful for enhanced sensitivity of TAPVS.

## 2. Device design and Modelling

The schematics in Fig. 1(a) displays the TAPVS with SS configuration. The long paths (brown) and side blocks (yellow) represents sensor/heater wires connecting pads respectively. The arrows (red) represents the acoustic signal inlet and outlet interacting with the heated wires. The cavity (purple) represents silicon wafer on which the device is fabricated. The cross sectional view in Fig. 1(b) provides detailed information about the material layers that are used for fabrication purpose. The top and bottom substrates made up of silicon representing the packaging cap and region below cavity respectively. The region between the silicon boundaries and multilayered wires is air cavity. The wires consists of three layers i.e. Platinum as heating material, chromium as adhesion layer and silicon nitride as mechanical support. The cross sectional view represents the key section of the device that is used for acquiring accurate simulation results. Fig.1(c) outlines the theoretical and computational foundation of SMTAPVS achieved through electromagnetic heating, one way coupling of thermo-viscous acoustics with joule heating and thermos-viscous acoustics with heat transfer in solids and fluids in time domain. This setup is implemented using Finite element method based software using electric currents (EC), Heat transfer in solids and fluids (HTS&F) and Thermo-viscous Acoustics (TVA) physics interfaces. The electromagnetic heating achieved using the coupling of EC and HTS&F is the basis of temperature rise in the heater. The EC interface solves a volumetric heat source where the relation between current density and the volumetric heat source is given by,

$$\nabla . J = Q_{J,V} \quad , \text{where} \quad J = \sigma E + J_e \quad \text{and} \quad E = -\nabla V \tag{1}$$

Here J is the current density, $Q_{J,V} = Q_e = J.E$ as the volumetric heat source, $\sigma$ is the electrical conductivity, E is the electric field, $J_e$ is the external current density, V is the electric scaler potential. The $Q_e$ is the coupling factor between EC and HTS&F. The HTS&F solves for the temperature using heat diffusion equation

$$\rho C_p u.\nabla T + \nabla . q = Q + Q_e \quad \text{with} \quad q = -k\nabla T \tag{2}$$

with temperature dependent material properties such as $\rho$ the electrical resistivity, $C_p$ the specific heat capacity at constant pressure, u the external medium velocity, T the temperature, q the conductive heat flux, k the thermal conductivity.

$$\rho C_p \frac{\partial T}{\partial t} + \rho C_p u.\nabla T = \nabla . (k\nabla T) + Q_e \tag{3}$$

The equation (3) is coupled equation between EC and HTS&F used for calculating the temperature rise in the

$$i\omega\rho_t + \nabla.(\rho_0 u_t) = 0 \tag{4}$$

where, equation (4) is continuity equation of mass conservation where $\omega$ is the acoustic frequency, $\rho_0$ is the air density, $\rho_t$ is the fluctuation in air density, $u_t$ is the total acoustic velocity field.

$$i\omega\rho_0 u_t = \nabla.\sigma \tag{5a}$$

$$\sigma = -p_t I + \mu(\nabla u_t + (\nabla u_t)^T) - \left(\frac{2}{3}\mu - \mu_B\right)(\nabla.u_t)I \tag{5b}$$

Equation (5a) is linearized Navier stokes equation of momentum conservation with $\sigma$ the stress tensor. Equation (5b) is constituting equation for stress tensor where, I is the identity tensor, $\mu$ is the dynamic viscosity, $\mu_B$ is the bulk viscosity.

$$\rho_0 C_p(i\omega T_t + u_t.\nabla T_0) - \alpha_p T_0(i\omega p_t + u_t.\nabla p_0) = \nabla.(k\nabla T_t) + Q \tag{6a}$$

$$\rho_t = \rho_0 (\beta_T p_t - \alpha_p T_t) \tag{6b}$$

$$p_t = p + p_b,\ u_t = u + u_b,\ T_t = T + T_b \tag{6c}$$

Equation (6a-c) is thermal equation of energy conservation where $T_t$ is the total temperature variation, $T_0$ is the equilibrium background temperature, $\alpha_p$ is the isobaric coefficient of thermal expansion, k is the temperature dependent thermal conductivity, $\beta_T$ is the isothermal compressibility. These equations incorporate the thermal and viscous losses to the heated wires due to the surrounding medium.

When acoustic flow interacts with the wires spaced at some distance, the thin boundary layers are developed around the boundaries. These boundary layers are named as thermal and viscous boundary layers which occurs due to thermal continuity and no slip conditions. Both thermal and viscous losses are comparable and depends on frequency as discussed below: -

$$\delta_v = \sqrt{\frac{\mu}{\pi f.\rho_0}} \tag{7}$$

where, $\delta_v$ is the Viscous boundary layer thickness, $\mu$ is the Dynamic viscosity of the medium, f is the Frequency of sound signal and $\rho_0$ is the Fluid density (kg/m$^3$).
In addition, thermal boundary layer is given as

$$\delta_t = \sqrt{\frac{k}{\pi f.\rho_0 C_p}} \tag{8}$$

where, $\delta_t$ is the Thermal boundary layer thickness, k is the Kinematic viscosity of the medium and $C_p$ is the Specific heat capacity at constant pressure.
The ratio of above two equations is described as

$$\frac{\delta_v}{\delta_t} = \sqrt{\frac{\mu C_p}{k}} = \sqrt{Pr} \tag{9}$$

where, Pr is the Prandt's Number

From above equations (7-8), it is clear that the thermal and viscous losses decrease with the increase in frequency. The implementation of boundary layers is crucial for designing and optimizing sensing wire. This setup ensures all relevant physics interfaces are used for accurate numerical simulations.

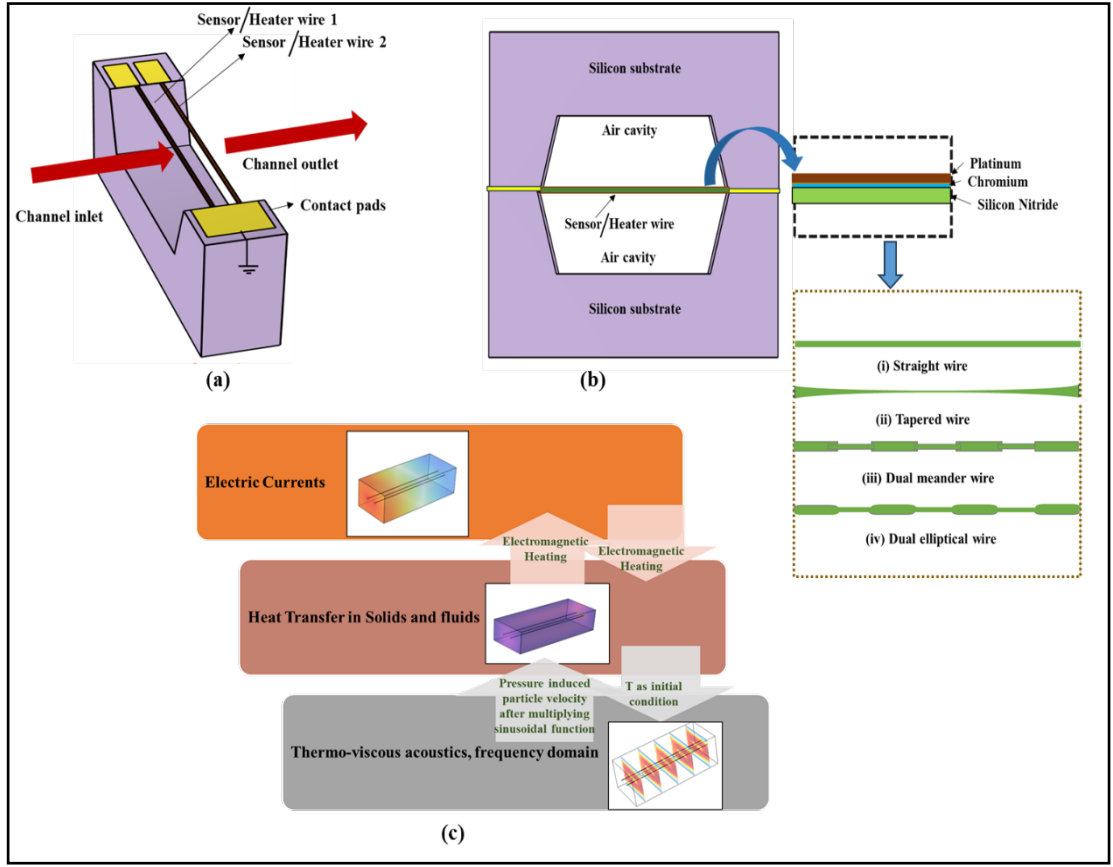

**Figure 1.** *Schematics for Structurally modified thermal particle velocity sensor structure (a) Schematic representation of TAPVS, with wires (brown), connecting pads (yellow), acoustic flow (red) (b) Cross sectional view of device structure used for numerical analysis and structural morphology of the wire as (i) Straight wire (ii) tapered wire (iii) Dual meander wire (iv) Dual elliptical wire (c) Theoretical framework of SMTAPVS incorporating Electric currents (EC), Heat transfer in solids and fluids(HTS&F) and thermos-viscous acoustics (TVA) physics interfaces*

In TAPVS, Both wires are heated to approximately 673K by supplying an equal magnitude of constant current to one terminal while grounding the common terminal of both wires. Heating creates a stationary and symmetric temperature distribution across the wires when no acoustic flow is present[33]. Thermal flow sensors facilitate the transfer of heat from high-temperature elements to low-temperature elements via the process of conduction, convection, and radiation. Heat transfer via radiation is minimal in comparison to convection; thus, it may be neglected[34][35][36]. Due to the absence of a direct connection between the hot wires, heat conduction is limited to interactions between the wires and the substrates. The reduction of this phenomenon is achieved through the application of an insulating layer, specifically $Si_3N_4$ or $SiO_2$. When acoustic flow interacts with the heated wires, it asymmetrically alters the temperature distribution along the direction of flow. The wires that come first in the direction of flow cool down more as compared to the latter one due to convection[6][33]. A barely noticeable temperature differential takes place between the wires, resulting in change in resistance. The variation in resistance can be converted into a voltage output by means of the utilization of the Wheatstone bridge circuit[37][38]. The relationship between particle vibration velocity along with the output voltage generated by the sound field has already been established.

$$U_{out} = \frac{U_0 \, \alpha \Delta T}{2} \qquad (10)$$

where $U_{out}$ is the differential output voltage, which is usually calculated after amplifying the signal, α represents the temperature coefficient of resistance, ΔT represents the variation in temperature between two wires and $U_0$ is voltage applied over both the sensors[31][39][40].

The analytical model of two wire Particle velocity sensor is given by[43]:-

$$\Delta T(f = 0) = e^{i2\pi f \frac{2v}{D/d}} \frac{P}{4\pi kL}\left(\ln\left(\frac{\pi d}{2L}\right) + \gamma\right) \tag{11}$$

where $\gamma$ is the Euler's constant having value approximately 0.577, f is the frequency of acoustic signal, v is input particle velocity, D is diffusion coefficient, d is separation between two wires, P is input wire, k is thermal conductivity of medium, and L is length of wire. By varying L and d in equation (11) respectively, the response of sensor for various dimensions is analysed. It is seen in Figure 3 that temperature difference is maximum using length as 1000 μm and wire separation as 110 μm dimensions. The sensor consists of two hot wires with identical dimensions 1000 ×2×0.1μm$^3$ dimensions with 110 μm distance between each wire[44]

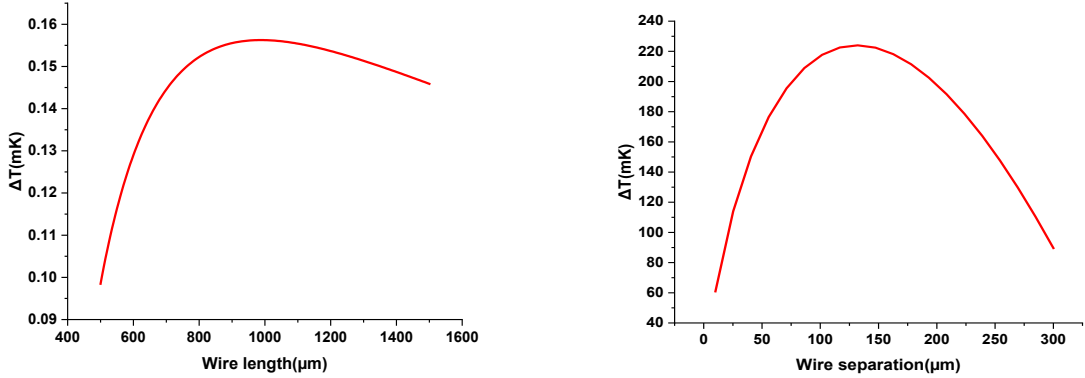

**Figure 2.** *The output temperature difference of TAPVS (a) Varying wire length (maximum at L = 1000 μm) (b) Varying wire separation (maximum at d = 110 μm)*

The microheater's material and design are essential for providing minimal thermal mass, low power consumption, efficient temperature homogeneity across the device, and superior thermal isolation from the environment[41][42].To achieve this, different structural modifications of hot microwire have been designed, such as (i) tapered wire (ii) dual meander wire and (iii) dual elliptical wire. The widths of SMTAPVS wires are varied in order to facilitate improved thermal isolation from its surroundings, reduced power usage, and low thermal mass[41][42]. Both dual meander and dual elliptical structures provide better temperature distribution of wire[42].The widths are varied in such a manner that overall volume of the identical wires remains approximately same. Structurally modified dimensions of SMTAPVS used in the present study are (i) Straight Wire( 1000μm × 2μm ×100nm) (ii) Tapered wire($W_1$ =1.785μm and $W_2$ = 2.785μm, L =1000μm , H = 100nm) (iii)Dual meander wire($N_1$= 41, $W_1$= 3μm & $N_2$ =40,$W_2$= 1μm, $L_1$=$L_2$ = 12.34 μm, H = 100nm) (iv) Dual elliptical wire($N_1$= 41, $W_1$= 3μm & $N_2$ =40,$W_2$= 1μm, $L_1$=$L_2$ = 12.34 μm, H = 100nm).

## 3. 3D Numerical Analysis

A rigorous three dimensional finite element method-based numerical analysis was utilized to create the computational model of SMTAPVS. In a nutshell, the stationary solution of a three-dimensional (3D) model is used as an initial condition to investigate the sensor response to the acoustic signal. It can be noted that only the sensing part of the SMTAPVS has been simulated to reduce computation cost. The sensing part consists of three layers: platinum as heating material, a $Si_3N_4$ layer to provide mechanical support to sensing wires, and chromium as an adhesion layer between the heater and insulating $Si_3N_4$. The standard properties of above discussed materials are used for numerical analysis.

For simulations to produce realistic results, temperature-dependent material characteristics like density, specific heat capacity, and thermal conductivity are essential. These properties tend to vary with temperature, which moderates the rate of temperature increase in SMTAPVS. Consequently, the sensitivity of the device will also reduce at higher heating power, which is in accord with the proposed finite element-based models[27][20][45].

### (i) Natural Frequency of SMTAPVS

In MEMS devices, it is necessary to calculate the natural frequency of it to avoid resonance at the device operating frequency. The Eigen frequencies for dual meander, dual elliptical, tapered and straight wire are 2.9069 kHz, 2.8665 kHz, 6.59 kHz and 5.83 kHz respectively. Therefore, Dual meander and Dual elliptical structures are only preferred for low frequencies i.e. < 1000Hz applications. On the other hand, Tapered and straight wires structures can be used for low frequencies i.e. 20Hz-2.5 KHz frequency applications

### (ii) Array optimization of Dual meander and Dual elliptical structured wire

Figure 3 illustrates the change in temperature of sensing wire with the increasing array size keeping constant current 2.93mA. The graph(red and grey) ensures the decrease in average temperature of sensing wire with array size sweep from 10 to 285. This is because the losses through the wire correspondingly increase, attributed to the rise in the surface to volume ratio[42].To achieve an optimal balance between minimizing losses and maintaining $T_{avg}$, an array size is chosen where further increases result in only a marginal decrease in $T_{avg}$. Based on the analysis, an array size of 40 is identified as the optimal value and is subsequently used for further simulation.

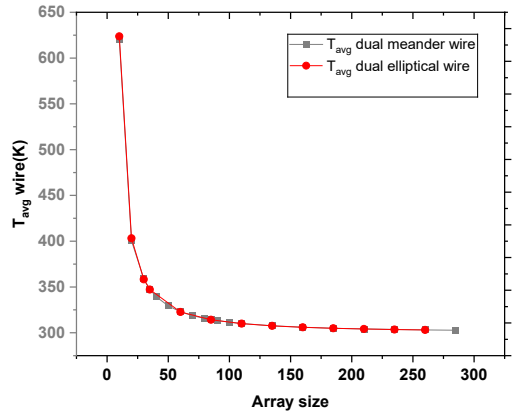

**Figure 3.** Array optimization of dual meander (grey) and dual elliptical (red) wire

The graph in Figure 4(a) compares the Joule heating performance of different SMAPVS wire structures: dual elliptical, dual meander, tapered, and straight. It is observed that the dual elliptical and dual meander structures (red and grey) consume less power, requiring an input current of approximately 2.93 mA to reach a temperature of ~673 K. In contrast, the tapered wire(green) requires 3.36 mA, while the conventional straight wire(blue) needs 3.45 mA to achieve the same temperature. It is because resistance dominates the narrower region of the wires in case of tapered wire making the overall resistance higher. In case of Dual meander and Dual elliptical structures, smaller blocks increase the resistance due to decrease in surface area of that region. These structures results in uneven current density as electrons gets crowded near the corners and edges of larger sized blocks. Due to increase in resistance, the required current flow decreases in the discussed structures. Figure 4(b) presents the cross-sectional view of temperature profile at the centre of sensing wires. It is observed that the temperature profile for tapered wire is maximum at the centre, succeeded by straight, dual meander and dual elliptical wire.

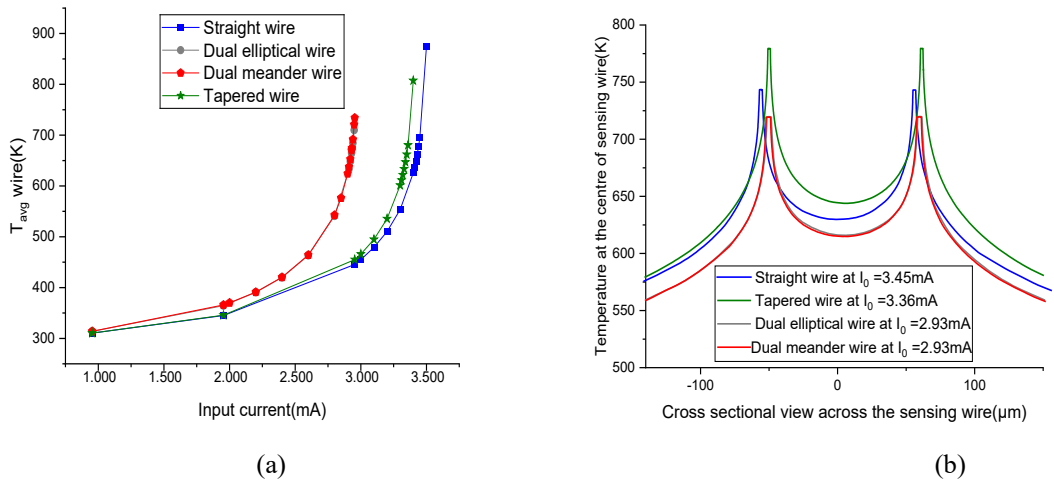

(a)                                                                 (b)

**Figure 4** *Electromagnetic heating of SMTAPVS structures* (a) constant current vs. rise in temperature of sensing wire for Dual elliptical (green), straight (blue), tapered (red) and dual meander (grey) wires (b) cross sectional temperature profile of heating wires at the centre of wires

Figure 5 shows the comparative time dependent electro-thermal heating of SMTAPVS structures. It is observed from the graph that straight wire will reach the steady state within 15ms time whereas dual elliptical, tapered and dual meander need 30ms, 15ms and 20ms to achieve the optimum temperature distribution. The SMTAPVS requires more time to achieve the optimum temperature due to thermal losses, thermal mass variation and non-linear current distribution due to varying cross sectional area long its length[42].

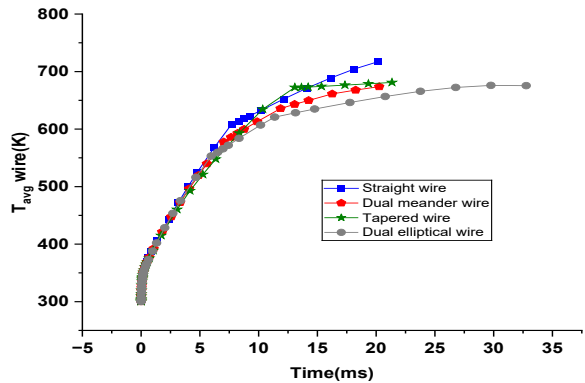

**Figure 5** Time dependent joule heating of SMTAPVS

In Figure 6, the comparative temperature difference is shown for each wire structure. It can be clearly seen that all the structurally modified structures show greater temperature difference compared to the conventional straight wire structures. The temperature difference is directly proportional to the particle velocity of the sound signal and is calculated by converting it into a voltage signal.

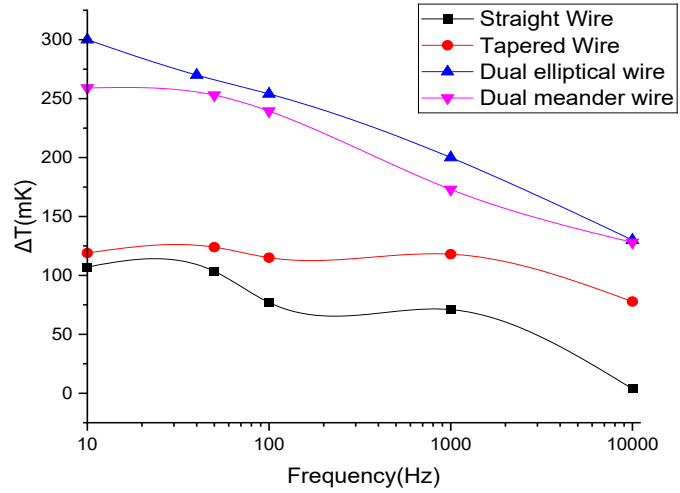

**Figure 6.** Comparative sensitivity of SMTAPVS

## 5. Conclusion

Three SMAPVS were analysed using FEM-based simulations to assess their potential for use in acoustic applications. The primary distinction among these sensor structures lies in the variation of their cross-sectional area along their length, which influences their performance characteristics. We first discussed the working principle of SMTAPVS and then established the numerical model using finite element method-based simulations. At steady state temperature, both sensor-heater wires are exactly at the same temperature. As the sound source induces particle motion in the vicinity of the SMTAPVS, a temperature difference is established across the two sensing wires. The overall temperature distribution along each wire with acoustic wave propagation, can be calculated by adding thermal perturbations to the steady-state temperature field. A complete 3D numerical model of SMTAPVS has been developed incorporating electromagnetic heating, Visco-thermal losses, and heat transfer modules. The electric currents and heat transfer module are simulated to optimize the temperature distribution of the wires by providing constant current to sensor wires. The corresponding particle velocity in the fluid medium is analysed using a Visco-thermal acoustic module, with a 1 Pa pressure applied at one end of the sensor. The resulting particle velocity and steady-state temperature are used as initial conditions for calculating the overall temperature difference between the wires. It is observed that all the structurally modified structures have higher temperature difference as compared to the conventional micro flown structures. These findings are fruitful for designing futuristic TAPVS having higher sensitivity.

## CrediT authorship contribution statement

**Ruby Jindal :** Modelling and Computational analysis, writing – original draft. **Sushil Kumar Singh**: Supervision. **Ravibabu Mulaveesala** : Supervision and analysis verification. **Jolly Xavier** : Conceptualization and model analysis, Methodology, Supervision , Writing & editing.

## Declaration of Competing Interest

The authors do not possess any financial interests that are in conflict.

## Acknowledgements

RJ gratefully acknowledges SSPL for the research fellowship.

Abbreviation – FEM (Finite Element method), SMTAPVS (Structurally Modified Thermal Acoustic Particle Velocity Sensor)

**References:-**

[1] Gordon S. Kino, *Acoustic waves: Devices, Imaging and analog signal processing*. Prentice-Hall, 1987.

[2] A. Nehorai and E. Paldi, "Acoustic Vector Sensor Array processing," 1994. doi: https://doi.org/10.1109/78.317869.

[3] H.-E. de Bree and J. W. Wind, "The acoustic vector sensor: a versatile battlefield acoustics sensor," in *Proceedings of SPIE*, SPIE, May 2011, pp. 80470C(1–8). doi: 10.1117/12.884681.

[4] "Microflown Technologies." Accessed: Jun. 20, 2025. [Online]. Available: https://www.microflown.com/

[5] H.-E. De Bree, "The Microflown: An Acoustic Particle Velocity Sensor," *Acoust. Aust.*, vol. 31, pp. 91–93, 2003.

[6] H E De Bree, "An overview of microflown technologies," in *Acta acustica united with Acustica*, 2003, pp. 1–11.

[7] M. Javaid, A. Haleem, S. Rab, R. Pratap Singh, and R. Suman, "Sensors for daily life: A review," *Sensors Int.*, vol. 2, no. 100121, pp. 1–10, Jan. 2021, doi: 10.1016/j.sintl.2021.100121.

[8] G. Varade, L. Bhamare, H. Mehta, and N. Mitra, "Recent Progress in MEMS Based Acoustic Vector Sensor for Underwater Applications," in *7th International Conference on Electrical, Electronics and Information Engineering*, Institute of Electrical and Electronics Engineers Inc., 2021, pp. 308–312. doi: 10.1109/ICEEIE52663.2021.9616884.

[9] T. G. H. Basten, H. E. De Bree, and E. H. G. Tijs, "Localization and tracking of aircraft with ground based 3D sound probes," 2007.

[10] T. B. Sadasivan, T. Basten, and H. E. de Bree, "Acoustic vector sensor based intensity measurements for passive localization of small aircraft," *J. Acoust. Soc. India*, vol. 36, no. 1, pp. 8–13, 2009, [Online]. Available: http://scholar.google.com/scholar?q=related:fv62PWuyAYoJ:scholar.google.com/&hl=en&num=30&as_sdt=0,5%5Cnpapers2://publication/uuid/2C111B7C-08E2-477D-A882-C0588A6A84C3

[11] E. H. G. Tijs, H. E. De Bree, T. G. H. Basten, and M. Nosko, "Non destructive and in situ acoustic testing of inhomogeneous materials," 2007.

[12] H.-E. De Bree and T. Basten, "2008 SAE BRASIL Noise and Vibration Conference E Low sound level source path contribution on a HVAC Low sound level source path contribution on a HVAC."

[13] E. Tijs and H. E. De Bree, "Mapping 3D sound intensity streamlines in a car interior," *SAE Int.*, 2009, doi: 10.4271/2009-01-2175.

[14] H.-E. de Bree and E. Tijs, "PU nosecone intensity measurements in a wind tunnel," 2009.

[15] H. D. H. De Bree, J. Wind, and S. Sadasivan, "Broad banded acoustic vector sensors for outdoor monitoring propeller driven aircraft," 2010. [Online]. Available: http://misuremeccaniche.it/2010_bordercontrol.pdf

[16] G.C.H.E.de Croon et al and E.H.G.Tijs, "Hear and Avoid for Micro Air Vehicles," 2016.

[17] D. F. Comesaña, E. Jansen, Y. R. Seco, and H. E. De Bree, "Acoustic multi-mission sensor (AMMS) system for illegal firework localisation in an urban environment," 2014.

[18] W. Q. Jing, D. F. Comesaña, and D. Perezcabo, "Sound source localisation using a single acoustic vector sensor and multichannel microphone phased arrays," in *INTERNOISE 2014*, 2014, pp. 1–8. doi: 10.13140/2.1.1020.8961.

[19] J. Kotus and B. Kostek, "Measurements and visualization of sound intensity around the human head in free field using acoustic vector sensor," 2015. doi: 10.17743/jaes.2015.0009.

[20] L. Benvenuti, A. Catania, P. Bruschi, and M. Piotto, "Modeling and optimization of directive acoustical particle velocity sensors for ultrasonic applications," *Sensors Actuators, A Phys.*, vol. 318, Feb. 2021, doi: 10.1016/j.sna.2020.112504.

[21] H.-E. de Bree, H. V Jansen, T. S. J Lammerink, G. J. M Krijnen, and M. Elwenspoek, "Bi-directional fast flow sensor with a large dynamic range," 1999.

[22] X. Zhang, D. Liu, X. Li, Y. Liu, Q. Tu, and Y. Zhou, "Particle Velocity Sensor and Its Application in Near-Field Noise Scanning Measurement," *J. Appl. Math. Phys.*, vol. 07, no. 11, pp. 2902–2908, 2019, doi: 10.4236/jamp.2019.711199.

[23] J. and D. W. and K. H. and R. R. and K. G. Van Honschoten, "Noise reduction in acoustic measurements with a particle velocity sensor by means of a cross-correlation technique," *Acta Acust. united with Acust.*, vol. 90, no. 2, pp. 349–355, 2004.

[24] J. W. Van Honschoten, D. R. Yntema, and R. J. Wiegerink, "An integrated 3D sound intensity sensor using four-wire particle velocity sensors: II. Modelling," *J. Micromechanics Microengineering*, vol. 20, no. 1, 2009, doi: 10.1088/0960-1317/20/1/015043.

[25] D. R. Yntema, J. W. Van Honschoten, and R. J. Wiegerink, "An integrated 3D sound intensity sensor using four-wire particle velocity sensors: I. Design and characterization," *J. Micromechanics Microengineering*, vol. 20, no. 1, Jan. 2010, doi: 10.1088/0960-1317/20/1/015042.

[26] O. Pjetri, R. J. Wiegerink, and G. J. M. Krijnen, "A 2D Particle Velocity Sensor with Minimal Flow Disturbance," *IEEE Sens. J.*, vol. 16, no. 24, pp. 8706–8714, 2016, doi: 10.1109/JSEN.2016.2570213.

[27] M. Piotto, F. Butti, E. Zanetti, A. Di Pancrazio, G. Iannaccone, and P. Bruschi, "Characterization and modeling of CMOS-compatible acoustical particle velocity sensors for applications requiring low supply voltages," *Sensors Actuators, A Phys.*, vol. 229, pp. 192–202, Jun. 2015, doi: 10.1016/j.sna.2015.03.018.

[28] S. Silvestri and E. Schena, "Micromachined flow sensors in biomedical applications," 2012. doi: 10.3390/mi3020225.

[29] Z. Li, W. Chang, C. Gao, and Y. Hao, "A novel five-wire micro anemometer with 3D directionality for low speed air flow detection and acoustic particle velocity detecting capability," *J. Micromechanics Microengineering*, vol. 28, no. 4, Feb. 2018, doi: 10.1088/1361-6439/aaac63.

[30] D. S. and P. B. Massimo Piotto, Andrea Ria, "Design and Characterisation of Acoustic Particle Velocity Sensors fabricated with a commercial post-CMOS MEMS process," in *International Conference on Solid-State Sensors, Actuators and Microsystems and Eurosensors XXXIII*, IEEE, 2019, pp. 1839–1842.

[31] H.-E. De Bree, P. Leussink, T. Korthorst, H. Jansen, T. S. J. Lammerink, and M. Elwenspoek, "The μ-flown: a novel device for measuring acoustic flows," *Sensors Actuators A*, vol. 54, pp. 552–557, 1996.

[32] F. Jacobsen and H.-E. De Bree, "The Microflown Particle Velocity Sensor," in *Handbook of Signal processing in acoustics*, Springer, 2008, pp. 1283–1291.

[33] S. Tulana *et al.*, "Design and simulation of Acoustic particle Velocity sensor," in *IWPSD*, 2017.

[34] J. T. W. Kuo, L. Yu, and E. Meng, "Micromachined thermal flow sensors-A review," 2012. doi: 10.3390/mi3030550.

[35] S. Wang, Z. Yi, M. Qin, and Q. A. Huang, "Modeling, Simulation, and Fabrication of a 2-D Anemometer Based on a Temperature-Balanced Mode," *IEEE Sens. J.*, vol. 19,

no. 13, pp. 4796–4803, Jul. 2019, doi: 10.1109/JSEN.2019.2902867.

[36] E. Taiedinejad, A. G. Kordlar, J. Koohsorkhi, and G. Sadeghian, "A four-wire micro anemometer in double cross shape with high mechanical stability for high sensitive air flow," *Microelectron. Eng.*, vol. 262, no. March, p. 111831, 2022, doi: 10.1016/j.mee.2022.111831.

[37] X. Chen and J. Shen, "Numerical analysis of mixing behaviors of two types of E-shape micromixers," *Int. J. Heat Mass Transf.*, vol. 106, pp. 593–600, Mar. 2017, doi: 10.1016/j.ijheatmasstransfer.2016.09.034.

[38] J. N. Shen, B. L. Liu, L. H. Zhu, M. H. Xu, Y. B. Zeng, and H. Guo, "Numerical analysis of the performance of microflown with SHS configuration," in *Proceedings of the 2017 Symposium on Piezoelectricity, Acoustic Waves, and Device Applications, SPAWDA 2017*, 2017, pp. 283–287. doi: 10.1109/SPAWDA.2017.8340341.

[39] V. B. Svetovoy and I. A. Winter, "Model of the µ-flown microphone," *Sensors and Actuators*, vol. 86, pp. 171–181, 2000, [Online]. Available: www.elsevier.nlrlocatersna

[40] J. W. van Honschoten, "Optimisation of a two wire thermal sensor for flow and sound measurements," vol. 00, no. x, pp. 523–526.

[41] Velmathi G and Mohan S, "Design, Electro-Thermal Simulation and Geometrical Optimization of Double Spiral Shaped Microheater on a Suspended Membrane for Gas Sensing."

[42] Z. E. Jeroish, K. S. Bhuvaneshwari, F. Samsuri, and V. Narayanamurthy, "Microheater: material, design, fabrication, temperature control, and applications—a role in COVID-19," *Biomed. Microdevices*, vol. 24, no. 1, Mar. 2022, doi: 10.1007/s10544-021-00595-8.

[43] J. W. van (Johannes W. Honschoten and Océ Facility Services), *Modelling and optimisation of the microflown*. s.n.], 2004.

[44] M. J. M. Hezemans and D. Kapolka, "Feasibility Study To Adapt the Microflown Vector Sensor for Underwater Use," 2012. [Online]. Available: http://hdl.handle.net/10945/27845

[45] Z. Zhu, W. Chen, L. Yang, and C. Gao, "An analytical model of three-hot-wire," 2023.